\documentclass[12pt]{article}
\usepackage{epsfig}
\begin{document}
%%%%%%%%%%%%%%%%%%%%%%%%%%%%%%%%%%%%%%%%%%%%%%%%%%%%%%%%%%%%%%%%%%%%%%%%%
\title{Near-threshold $J/\psi$ photoproduction off nuclei}
\author{E. Ya. Paryev$^{1,2}$ and Yu. T. Kiselev$^2$\\
{\it $^1$Institute for Nuclear Research, Russian Academy of Sciences,}\\
{\it Moscow 117312, Russia}\\
{\it $^2$Institute for Theoretical and Experimental Physics,}\\
{\it Moscow 117218, Russia}}
%==============================================================
%%==============================================================

\renewcommand{\today}{}
\maketitle

\begin{abstract}
   We study the $J/\psi$ photoproduction from nuclei near the kinematic threshold within
   the first collision model, based on the nuclear spectral function,
   for incoherent primary photon--nucleon charmonium creation processes. The model takes into account the final $J/\psi$ absorption, target nucleon binding and Fermi motion, the formation length of $J/\psi$ mesons as well as the effect of their nuclear mean-field potential on these processes. We calculate the A dependences of the absolute and
   relative (transparency ratio) charmonium yields as well as its absolute and relative
   excitation functions within the different
   scenarios for the $J/\psi N$ absorption cross section, for the $J/\psi$ formation length and for $J/\psi$ in-medium
   modification. We demonstrate that the studied observables, on the one hand, are not practically affected
   by the charmonium formation length and mass shift
   effects and, on the other hand, they are appreciably sensitive to the genuine
   $J/\psi N$ absorption cross section at above threshold beam energies, which means that these observables can
   be useful to help determine the $J/\psi N$ absorption cross section from the comparison of the results of our calculations with the future data from the experiments in the Hall C at the upgraded up to 12 GeV  CEBAF facility.
   We also show that the absolute and relative excitation functions for $J/\psi$ subthreshold production in ${\gamma}A$ reactions reveal some sensitivity to adopted in-medium modification scenarios for $J/\psi$ mesons.
   Therefore, such observables, measured in the subthreshold energy domain, may be an important tool to get valuable information on the charmonium in-medium properties in cold nuclear matter.
\end{abstract}

\newpage

\section*{1. Introduction}

\hspace{1.5cm} The production and suppression of $J/\psi$ mesons on a nuclear targets have been intensively
studied over the last thirty years, both experimentally and theoretically. Most investigations of these
issues have been carried out in the field of relativistic proton--nucleus and nucleus--nucleus collisions
(see, e.g., [1--28]). The main goal of the studies in this field was to get valuable information about
a possible restoration of chiral symmetry above some critical temperature and density accompanied by the
phase transition from composite hadrons to a quark-gluon plasma.
And the suppression of $J/\psi$ yield observed in heavy--ion collisions at SPS and RHIC
is recognized, beginning from the work [29], as one of the most promising signals of the
transition of ordinary hadronic matter to a deconfined phase of quarks and gluons. In addition to the above
prominent prospect, in-medium properties of charmonium states, the strength of the inelastic $J/\psi$--nucleon interaction are of further interest in view both of the feasibility of an experimental observation of the one
of the new kinds of exotic nuclei--narrow bound $J/\psi$--nucleus states [30] and of gaining more insight into
the dynamics of low-energy QCD in the charm and hidden charm sectors [31, 32]. What concerns the experimental
situation, up to now a wide variety of high statistics $J/\psi$ data samples have been collected in
proton--nucleus and heavy--ion experiments NA38 [6], NA50 [7] and NA60 [8] at the CERN-SPS, in PHENIX and
STAR experiments [9] at RHIC as well as in recent ALICE experiments [10] at the CERN-LHC. Moreover, nowadays
there are charmonium results in $pA$ collisions at initial proton beam energy of 920 GeV from HERA-B experiment
at the HERA proton ring of the DESY laboratory [11] and from E866 experiment at FNAL at beam energy of 800 GeV
[12]. Various theoretical approaches have been proposed (see, e.g., [1--5, 13--28]) for the description of
$J/\psi$ production in $pA$ and $AA$ interactions in the SPS, RHIC and LHC energy domains. The current theoretical
and experimental status of the study of charmonium production in high-energy nuclear collisions is given in [33].

   Due to almost negligible initial state interactions, the photon--nucleus reactions represent also an important
and more transparent compared to proton--nucleus and heavy--ion collisions tool to investigate the charmonium
in-medium properties, its interaction in cold nuclear matter. Several $J/\psi$ production experiments, using
nuclear targets, have been performed at high photon energies, starting with 20 GeV (see Refs. [34--40]).
The existing data on $J/\psi$ creation off only two nuclei (d, Be) below 20 GeV come from Cornell [41] using
9.3--11.8 GeV photons and from SLAC [42] from 13--21 GeV photons. No $J/\psi$ production events were observed
in recent subthreshold
\footnote{$^)$The threshold energy for $J/\psi$ photoproduction on a free nucleon being at rest is 8.2 GeV.}$^)$
JLab experiment [43], which used a carbon target and photons with energies below 6 GeV. So, it is not much is
known presently in the region of the photon energies around the threshold energy of 8.2 GeV. All above calls for a
new measurements in this region of photon energies. Various theoretical aspects of $J/\psi$ photoproduction on
nuclei at high energies are discussed, e.g., in [44--51].

  A particular interest was shown in recent years in the study of the charmonium production and suppression
in proton--nucleus [52--54] and antiproton--nucleus [55--60] as well as in photon--nucleus [61, 62] reactions
at energies close to the kinematic thresholds on free nucleons. This interest was triggered by the fact that
these reactions are planned to be investigated in the near future in the CBM and ${\rm {\bar P}}$ANDA
experiments at the constructed FAIR facility in Germany [63] as well as at the upgraded up to 12 GeV CEBAF
accelerator in the USA [64].
Since direct measurement of the fundamental $J/\psi$--nucleon inelastic cross section is not possible,
analysis of data, collected in these experiments, will allow to perform a much improved determination of it
because in low-energy collisions, contrary to the high-energy ones,
the $J/\psi$ mesons are produced with low momenta relative to the nuclear medium at which their formation
time (or color transparency) effects are expected to be insignificant and, therefore, the situation with
the scattering of a full-sized charmonium (but not a compact $c\bar{c}$-pair) on an intranuclear nucleons
with the full $J/\psi$ meson--nucleon cross section will be realized.

The primary goal of the present work is to extend the first collision model based on
the nuclear spectral function [54] that has been adopted by us for the description of the A dependences
of the absolute and relative (transparency ratio) $J/\psi$ meson yields as well as its excitation function
in $pA$ collisions near threshold to $J/\psi$--producing electromagnetic processes. In this paper we present
the estimates for the above observables from ${\gamma}A$ reactions in the near-threshold energy region
obtained within this extended model. In view of the expected data from the JLab upgraded to 12 GeV,
these estimates can be used as an important tool for determining the genuine
$J/\psi$--nucleon absorption cross section as well as for understanding the properties of charmonium
in nuclear medium and the role played by the various nucleus-related effects in the $J/\psi$
production near threshold.

\section*{2. Framework: a direct knock out processes}

\hspace{1.5cm} Since we are interested in the near-threshold incident photon energy region up to
11 GeV, we have taken into consideration the following elementary processes,
which have the lowest free production threshold ($\approx$ 8.2 GeV)
\footnote{$^)$We can neglect in the energy domain of our interest the following two-step $J/\psi$ production
processes with $\chi_{c1}$, $\chi_{c2}$ and $\Psi^\prime$ mesons in an intermediate states:
 ${\gamma}N \to \chi_{c1}N$, ${\gamma}N \to \chi_{c2}N$, ${\gamma}N \to {\Psi^\prime}N$;
 $\chi_{c1} \to J/\psi\gamma$
 (${\rm BR}=36$\%), $\chi_{c2} \to J/\psi\gamma$ (${\rm BR}=20$\%),
 $\Psi^\prime \to J/\psi{\pi}{\pi}$ (${\rm BR}=49$\%) and $\Psi^\prime N \to J/\psi N$ due to larger
$\chi_{c1}$, $\chi_{c2}$ and $\Psi^\prime$ production thresholds in ${\gamma}N$ collisions--10.1, 10.3 and
10.9 GeV, respectively.}$^)$
:
%formula(1)
\begin{equation}
{\gamma}+p \to J/\psi+p,
\end{equation}
%formula(2)
\begin{equation}
{\gamma}+n \to J/\psi+n.
\end{equation}
In line with [54], in the following calculations we will include
the medium modification of the $J/\psi$ mesons, participating in the production processes (1),
(2), by using, for reasons of simplicity, their average in-medium mass $<m^*_{{J/\psi}}>$ defined as:
%formula(3)
\begin{equation}
<m^*_{{J/\psi}}>=\int d^3r{\rho_N({\bf r})}m^*_{{J/\psi}}({\bf r})/A,
\end{equation}
where ${\rho_N({\bf r})}$ and $m^*_{{J/\psi}}({\bf r})$ are the local nucleon density and
${J/\psi}$ effective mass inside the nucleus, respectively. Assuming that
%formula(4)
\begin{equation}
m^*_{{J/\psi}}({\bf r})=m_{{J/\psi}}+V_0\frac{{\rho_N({\bf r})}}{{\rho_0}},
\end{equation}
we can readily rewrite equation (3) in the form
%formula(5)
\begin{equation}
<m^*_{{J/\psi}}>=m_{{J/\psi}}+V_0\frac{<{\rho_N}>}{{\rho_0}}.
\end{equation}
Here, $m_{{J/\psi}}$ is the ${J/\psi}$ free space mass, $<{\rho_N}>$ and ${\rho_0}=0.16$ fm$^{-3}$ are the
average and saturation nucleon densities, respectively.
Our calculations show that, for example, for target nuclei $^{12}$C, $^{40}$Ca,
$^{93}$Nb and $^{208}$Pb the ratio $<{\rho_N}>/{\rho_0}$ is approximately equal to 0.5, 0.6, 0.7 and 0.8, respectively.
We will use the respective above values throughout the following study. In it in line with [54]
for the $J/\psi$ mass shift at
saturation density $V_0$ we will employ the five following options: i) $V_0=0$, ii) $V_0=-25$ MeV,
iii) $V_0=-50$ MeV, iv) $V_0=-100$ MeV, and v) $V_0=-150$ MeV.
It should be pointed out that according to the theory expectations [65--70] the $J/\psi$ mass shift at
normal nuclear matter density is either very small or $\sim$ -20 MeV due to a weak coupling of the $c$, ${\bar c}$
quarks to the nuclear medium. However, in reality it is unclear presently which shift is the correct one.
Therefore, to extend the potential range of applicability of our model we will perform the calculations
of the $J/\psi$ photoproduction cross sections off nuclei in the scenarios with possible charmonium
mass shift $V_0$ ranging from -150 MeV to 0.
As in [54], we will ignore the medium modification of the outgoing nucleon mass in the present work.

  Then, ignoring the distortion of the incident photon in the energy range of our interest
and using the results given in [54],
we can represent the total cross section for the production of ${J/\psi}$ mesons
off nuclei in the primary photon--induced reaction channels (1), (2) as follows:
%formula(6)
\begin{equation}
\sigma_{{\gamma}A\to {J/\psi}X}^{({\rm prim})}(E_{\gamma})=I_{V}[A]
\left<\sigma_{{\gamma}N \to {J/\psi}N}(E_{\gamma})\right>_A,
\end{equation}
where
%formula(7)
\begin{equation}
I_{V}[A]=2{\pi}A\int\limits_{0}^{R}r_{\bot}dr_{\bot}
\int\limits_{-\sqrt{R^2-r_{\bot}^2}}^{\sqrt{R^2-r_{\bot}^2}}dz
\rho(\sqrt{r_{\bot}^2+z^2})
\exp{\left[-A\int\limits_{z}^{\sqrt{R^2-r_{\bot}^2}}\sigma_{{J/\psi}N}^{\rm eff}(x-z)
\rho(\sqrt{r_{\bot}^2+x^2})dx\right]},
\end{equation}
%formula(8)
\begin{equation}
\left<\sigma_{{\gamma}N \to {J/\psi}N}(E_{\gamma})\right>_A=
\int\int
P_A({\bf p}_t,E)d{\bf p}_tdE
\sigma_{{\gamma}N \to {J/\psi}N}(\sqrt{s},<m^*_{{J/\psi}}>)
\end{equation}
and
%formula(9)
\begin{equation}
  s=(E_{\gamma}+E_t)^2-({\bf p}_{\gamma}+{\bf p}_t)^2,
\end{equation}
%formula(10)
\begin{equation}
   E_t=M_A-\sqrt{(-{\bf p}_t)^2+(M_{A}-m_{N}+E)^{2}}.
\end{equation}
Here,
$\sigma_{{\gamma}N\to {J/\psi}N}(\sqrt{s},<m^*_{{J/\psi}}>)$ is the "in-medium"
total cross section for the production of ${J/\psi}$
with reduced mass $<m^*_{{J/\psi}}>$ in reactions (1) and (2)
\footnote{$^)$In equation (6) it is assumed that the $J/\psi$ meson production cross sections
in ${\gamma}p$ and ${\gamma}n$ interactions are the same [42].}$^)$
at the ${\gamma}N$ center-of-mass energy $\sqrt{s}$;
$\rho({\bf r})$ and $P_A({\bf p}_t,E)$ are the local nucleon density and the
spectral function of target nucleus $A$ normalized to unity
\footnote{$^)$The specific information about these quantities, used in our subsequent calculations,
is given in [71, 72].}$^)$;
${\bf p}_{t}$  and $E$ are the internal momentum and binding energy of the struck target nucleon
just before the collision; $A$ is the number of nucleons in
the target nucleus, $M_{A}$  and $R$ are its mass and radius; $m_N$ is the bare nucleon mass;
${\bf p}_{\gamma}$ and $E_{\gamma}$ are the momentum and energy of the initial photon;
$\sigma_{{J/\psi}N}^{\rm eff}(z)$ is the $J/\psi$--nucleon effective absorption cross section, which
will be defined below. The quantity $I_{V}[A]$ in equation (6) represents the effective number of target
nucleons participating in the primary ${\gamma}N \to {J/\psi}N$ reactions. It should be noticed that the
expression (7) for it is valid only at low photon energies at which the so-called coherence length $l_c$--
the distance that the virtual $c\bar{c}$ fluctuation of the incoming photon travels in the lab frame before
scattering elastically on the nucleon--is much less than the nucleus size [59]. Accounting for that the
coherence length $l_c=2E_{\gamma}/m_{J/\psi}^2$ [59], we get that at the photon energy $E_{\gamma}=11$ GeV,
accessible at the JLab upgraded to 12 GeV, it is about of 0.5 fm. This is substantially less than the nucleus
radius, which is to say that in the near-threshold $J/\psi$ photoproduction on nuclei the $c\bar{c}$-pair is
created essentially right at the location of the nucleon that it scatters from. After this scattering the
compact $c\bar{c}$ state evolves over some time (the formation time) or over some distance (the formation
length, see also below) to form the final full-sized $J/\psi$ meson.

  Following [54], we assume that the "in-medium" cross section
$\sigma_{{\gamma}N \to {J/\psi}N}({\sqrt{s}},<m^*_{J/\psi}>)$ for $J/\psi$ production in reactions (1) and (2)
is equivalent to the vacuum cross section $\sigma_{{\gamma}N \to {J/\psi}N}({\sqrt{s}},m_{J/\psi})$ in which
the free mass $m_{J/\psi}$ is replaced by the average in-medium mass $<m^*_{{J/\psi}}>$ as given by
equation (5). For the free total cross section $\sigma_{{\gamma}N \to {J/\psi}N}({\sqrt{s}},m_{J/\psi})$
in the photon energy range $E_{\gamma} \le $ 22 GeV we have used the following parametrization,
based on the near-threshold predictions of the two gluon exchange model [61]:
%formula(11)
\begin{equation}
\sigma_{{\gamma}N \to {J/\psi}N}({\sqrt{s}},m_{J/\psi})=11.1(1-x)^2~[{\rm nb}],
\end{equation}
where
\footnote{$^)$It is easily seen that also $x=E^{\rm thr}_{\gamma}/E_{\gamma}$,
where $E^{\rm thr}_{\gamma}=(s_{\rm thr}-m^2_N)/2m_N$ is the energy at kinematic threshold.}$^)$
%formula(12)
\begin{equation}
  x=(s_{\rm thr}-m^2_N)/(s-m^2_N), \,\,\,\,s_{\rm thr}=(m_{J/\psi}+m_N)^2, \,\,\,\,
  s=(E_{\gamma}+m_N)^2-{\bf p}_{\gamma}^2.
\end{equation}
The comparison of the results of calculations by (11) (solid line) with the scarce existing data from
"Cornell 75" [41] (full dot), "SLAC 75" [42] (full triangles) and "SLAC 76 unpublished" [73] (open triangles),
collected together in [64], is shown in figure 1. It can be seen that the parametrization (11) fits well the
existing set of data for the ${\gamma}N \to {J/\psi}N$ reaction in the considered range of photon energies.
%%%%%%%%%%%%%%%%%%%%%%%%%%%%%%%%%%%%%%%%%%%%%%%%%%%%%%%%%%%
\begin{figure}[htb]
\begin{center}
\includegraphics[width=12.0cm]{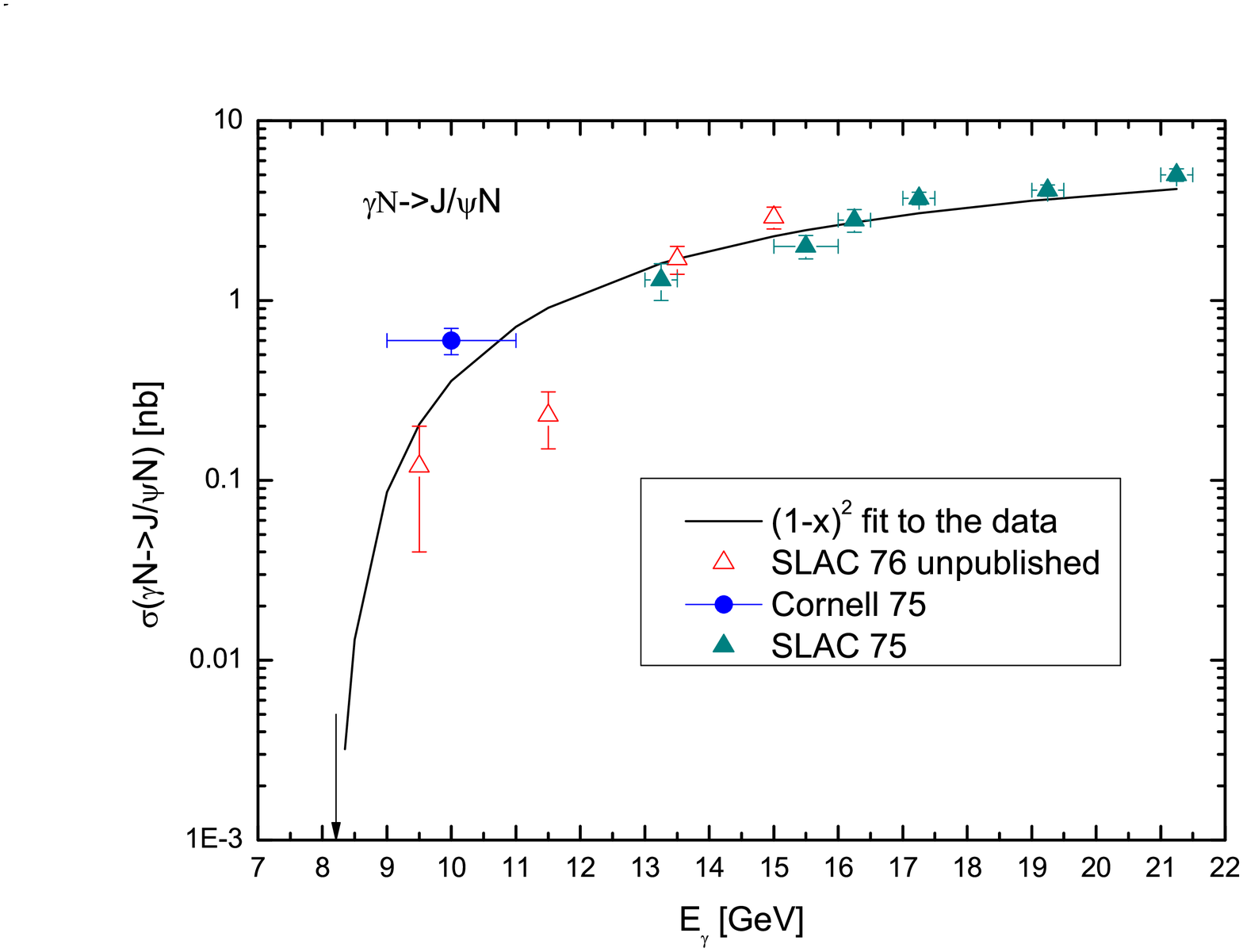}
\vspace*{-2mm} \caption{(color online) Total cross section for the reaction ${\gamma}N \to {J/\psi}N$ as a function
of photon energy. The arrow indicates its threshold on a free nucleon. For notation see the text.}
\label{void}
\end{center}
\end{figure}
%%%%%%%%%%%%%%%%%%%%%%%%%%%%%%%%%%%%%%%%%%%%%%%%%%%%%%%%%%%%%%%%%%%%%%%%%%%%%%%%%%%%%%%%%%%%%

 Following [54, 56, 59, 74], we express the charmonium--nucleon effective cross section
$\sigma_{{J/\psi}N}^{\rm eff}(z)$, entering into the equation (7) and taking into account the
time dependence of the $J/\psi$ formation, in terms of a $J/\psi$ meson formation length $l_{J/\psi}$:
%formula(13)
\begin{equation}
\sigma_{{J/\psi}N}^{\rm eff}(z)=\sigma_{{J/\psi}N}\left\{\theta(l_{J/\psi}-z)\left[\frac{z}{l_{J/\psi}}+
\frac{n^2<k^2_t>}{m^2_{J/\psi}}\left(1-\frac{z}{l_{J/\psi}}\right)\right]+\theta(z-l_{J/\psi})\right\}.
\end{equation}
Here, $\sigma_{{J/\psi}N}$ is the genuine free-space $J/\psi$--nucleon absorption cross section;
n is the number of valence quarks of the hadron ($n=2$ in our case), while $<k^2_t>^{1/2}$ is the
average transverse momentum of the quark in the hadron (taken to be $<k^2_t>^{1/2}=0.35$ GeV/c);
$z$ is the distance from the production point of the $c\bar{c}$-pair which evolves over the formation
length $l_{J/\psi}$ into the final $J/\psi$ meson and $\theta(x)$ is the standard step function.
For the $J/\psi$ meson formation length $l_{J/\psi}$ we adopt the conventional formula with an energy
denominator [54, 56, 59, 74]:
%formula(14)
\begin{equation}
l_{J/\psi}\simeq\frac{2p^{\rm lab}_{J/\psi}}{m^2_{\Psi^\prime}-m^2_{J/\psi}}
\approx 0.1~{\rm fm}~p^{\rm lab}_{J/\psi}/({\rm GeV}/{\rm c}),
\end{equation}
where $p^{\rm lab}_{J/\psi}$ is the charmonium momentum in the target nucleus rest frame.
Taking into consideration that the average momentum between the maximum and minimum allowable
momenta in the process ${\gamma}N \to {J/\psi}N$ proceeding on a free target nucleon being
at rest at initial photon energy of 11 GeV is $p^{\rm lab}_{J/\psi}\simeq{7.7}$ GeV/c,
one can get that $l_{J/\psi}\simeq{0.8}$ fm for this momentum. We will use this value for the
quantity $l_{J/\psi}$ throughout our calculations for a collection of $A$ target nucleons subject to
Fermi motion in the near-threshold energy domain.
 Ignoring the struck target nucleon binding and Fermi motion and outgoing $J/\psi$ in-medium modifications, from (6)
we get the following simple expression for the total cross section
$\sigma_{{\gamma}A\to {J/\psi}X}^{({\rm prim})}(E_{\gamma})$:
%formula(15)
\begin{equation}
\sigma_{{\gamma}A\to {J/\psi}X}^{({\rm prim})}(E_{\gamma})=I_{V}[A]
\sigma_{{\gamma}N \to {J/\psi}N}(\sqrt{s},m_{J/\psi}),
\end{equation}
where the elementary cross section $\sigma_{{\gamma}N \to {J/\psi}N}(\sqrt{s},m_{J/\psi})$ is given by (11).
It is valid only in a photon energy range above threshold (see figures 2, 5 and 6, given below) and allows one
to easily estimate here this cross section.

     The absorption cross section $\sigma_{{J/\psi}N}$ can be extracted, in particular, from
a comparison of the calculations with the measured transparency ratio of the $J/\psi$ meson,
normalized, for example, to carbon:
%formula(16)
\begin{equation}
T_A=\frac{12~\sigma_{{\gamma}A \to {J/\psi}X}(E_{\gamma})}{A~\sigma_{{\gamma}{\rm C} \to {J/\psi}X}(E_{\gamma})}.
\end{equation}
Here, $\sigma_{{\gamma}A \to {J/\psi}X}(E_{\gamma})$ and $\sigma_{{\gamma}{\rm C} \to {J/\psi}X}(E_{\gamma})$
are inclusive total cross sections for $J/\psi$ production in
${\gamma}A$ and ${\gamma}{\rm C}$ collisions at incident photon energy $E_{\gamma}$, respectively.
If the primary photon--induced reaction channels
(1), (2) dominate in the $J/\psi$ production in ${\gamma}A$ reactions close to threshold
\footnote{$^)$One may expect that this is so due to the following. The main inelastic channel in ${\gamma}N$
collisions at beam energies of interest is the multiplicity production of pions with comparatively
low energies at which the secondary ${\pi}N \to {J/\psi}X$ processes are energetically suppressed.}
$^)$
, then, according to (6) and (7), we have:
%formula(17)
\begin{equation}
T_A=\frac{12~\sigma_{{\gamma}A \to {J/\psi}X}^{({\rm prim})}(E_{\gamma})}{A~\sigma_{{\gamma}{\rm C} \to {J/\psi}X}^{({\rm prim})}(E_{\gamma})}=
\frac{12~I_V[A]}{A~I_V[{\rm C}]}
\frac{\left<\sigma_{{\gamma}N \to {J/\psi}N}(E_{\gamma})\right>_A}
{\left<\sigma_{{\gamma}N \to {J/\psi}N}(E_{\gamma})\right>_{\rm C}}.
\end{equation}
Ignoring the difference between the cross sections $\left<\sigma_{{\gamma}N \to {J/\psi}N}(E_{\gamma})\right>_A$
and $\left<\sigma_{{\gamma}N \to {J/\psi}N}(E_{\gamma})\right>_{\rm C}$
\footnote{$^)$This difference is small (within several percent) both at above threshold
incident energies and at energies just below the threshold, as our calculations by (17) and (18)
for the nucleus with a diffuse boundary showed. However, it becomes substantial at far subthreshold beam energies
as our calculations also demonstrated (compare, for example, short-dashed and solid lines in figure 4
given below). It is easily seen that, according to equations (15), (17), the expression (18)
describes the transparency ratio $T_A$ also in the case when primary ${\gamma}N \to {J/\psi}N$ processes
proceed on a free target nucleon being at rest without $J/\psi$ in-medium modifications.}$^)$,
from (17) we approximately obtain:
%formula(18)
\begin{equation}
T_A \approx \frac{12~I_V[A]}{A~I_V[{\rm C}]}.
\end{equation}
As is easy to see from (13), when $J/\psi$ formation length $l_{J/\psi} \to 0$ then
the exponent in equation (7) can be put in the following in an easy-to-use in analytical integration form:
%formula(19)
\begin{equation}
A\sigma_{{J/\psi}N}\int\limits_{z}^{\sqrt{R^2-r_{\bot}^2}}\rho(\sqrt{r_{\bot}^2+x^2})dx.
\end{equation}
In this case the integral (7) for the quantity $I_V[A]$ can be transformed to a simpler expression:
%formula(20)
\begin{equation}
I_{V}[A]=\frac{\pi}{\sigma_{{J/\psi}N}}\int\limits_{0}^{R^2}dr_{\bot}^2
\left(1-e^{-A\sigma_{{J/\psi}N}\int\limits_{-\sqrt{R^2-r_{\bot}^2}}^{\sqrt{R^2-r_{\bot}^2}}
\rho(\sqrt{r_{\bot}^2+x^2})dx}\right),
\end{equation}
which in the cases of Gaussian nuclear density ($\rho({\bf r})=(b/\pi)^{3/2}\exp{(-br^2)}$)
and a uniform nucleon density for a
nucleus of a radius $R=r_0A^{1/3}$ with a sharp boundary is reduced to even more simple forms:
%formula(21)
\begin{equation}
I_V[A]=\frac{A}{x_G}\int\limits_{0}^{1}\frac{dt}{t}\left(1-e^{-x_Gt}\right), \,\,\,\,x_G=A\sigma_{{J/\psi}N}b/\pi
\end{equation}
and
%formula(22)
\begin{equation}
I_V[A]=\frac{3A}{2a_1}\left\{1-\frac{2}{a_1^2}[1-(1+a_1)e^{-a_1}]\right\}, \,\,\,\,a_1=3A\sigma_{J/\psi N}/2{\pi}R^2,
\end{equation}
respectively.
The simple formulas (15), (21), (22) and (18), (21), (22) allow one to easily estimate both the total cross section $\sigma_{{\gamma}A\to {J/\psi}X}^{({\rm prim})}(E_{\gamma})$ at above threshold energies and the transparency ratio $T_A$ here as well as at initial energies not far below the threshold (see figures 3 and 4 given below).

  Let us discuss now the results of our calculations in the framework of the approach outlined above.

\section*{3. Results and discussion}

\hspace{1.5cm} Figure 2 shows the A--dependence of the total $J/\psi$ production
cross section from the primary ${\gamma}N \to {J/\psi}N$ reaction channels in ${\gamma}A$
($A=$$^{12}$C, $^{27}$Al, $^{40}$Ca, $^{93}$Nb, $^{208}$Pb, and $^{238}$U) collisions calculated
for incident photon energy of $E_{\gamma}=11$ GeV on the basis of equations (6) and (15) for the same
values of the genuine charmonium--nucleon absorption cross section $\sigma_{J/\psi N}$,
as those used in [54] and indicated in the inset, and for no $J/\psi$ mass shift.
As in [54], the calculations with only $l_{J/\psi}=0$ in equation (7)
are given in the figure for $\sigma_{J/\psi N}=7$ mb and $\sigma_{J/\psi N}=14$ mb, for the case of
$\sigma_{J/\psi N}=3.5$ mb the ones are presented here already for two options, namely: i) $l_{J/\psi}=0$
and ii) $l_{J/\psi}=0.8$ fm.
%%%%%%%%%%%%%%%%%%%%%%%%%%%%%%%%%%%%%%%%%%%%%%%%%%%%%%%%%%%
\begin{figure}[htb]
\begin{center}
\includegraphics[width=12.0cm]{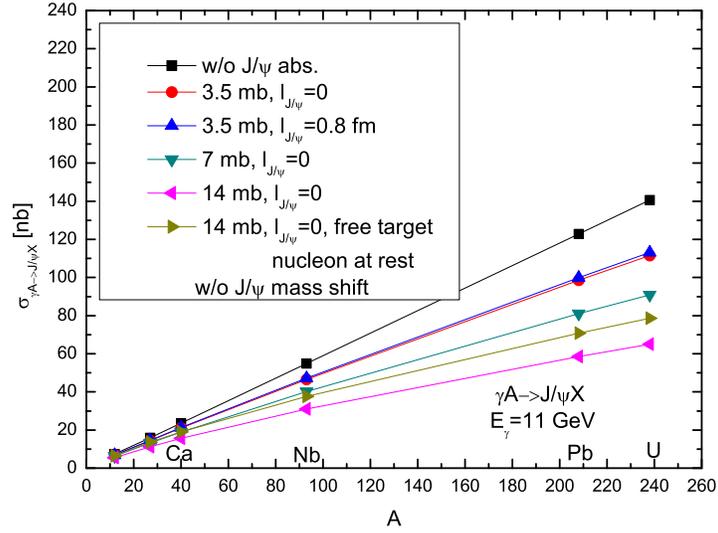}
\vspace*{-2mm} \caption{(color online) A--dependence of the total cross section of $J/\psi$ production
by 11 GeV photons in the full phase space from primary ${\gamma}N \to {J/\psi}N$ reactions
proceeding on an off-shell target nucleons and on a free ones being at rest in the scenario without $J/\psi$
mass shift for different values of the $J/\psi N$ absorption cross section and $J/\psi$ formation length indicated
in the inset. The lines are to guide the eyes.}
\label{void}
\end{center}
\end{figure}
%%%%%%%%%%%%%%%%%%%%%%%%%%%%%%%%%%%%%%%%%%%%%%%%%%%%%%%%%%%%%%%%%%%%%%%%%%%%%%%%%%%%%%%%%%%%%
%%%%%%%%%%%%%%%%%%%%%%%%%%%%%%%%%%%%%%%%%%%%%%%%%%%%%%%%%%%
\begin{figure}[!h]
\begin{center}
\includegraphics[width=12.0cm]{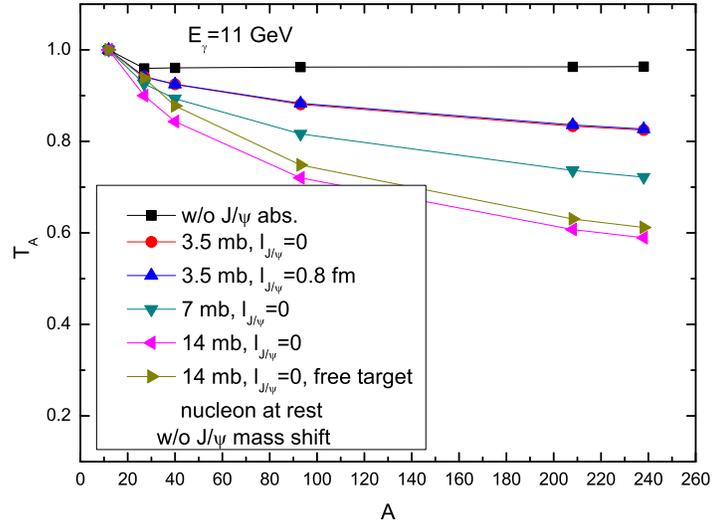}
\vspace*{-2mm} \caption{(color online) Transparency ratio $T_A$ for $J/\psi$ mesons
from primary ${\gamma}N \to {J/\psi}N$ reactions proceeding on an off-shell target nucleons
and on a free ones being at rest as a function of the nuclear mass number $A$
in the scenario without their mass shift as well as for their different absorption
cross sections and formation lengths indicated in the inset. The lines are to guide the eyes.}
\label{void}
\end{center}
\end{figure}
%%%%%%%%%%%%%%%%%%%%%%%%%%%%%%%%%%%%%%%%%%%%%%%%%%%%%%%%%%%%%%%%%%%%%%%%%%%%%%%%%%%%%%%%%%%%%
One can see that, as in the case of $pA$ reactions [54], the results are practically insensitive to the $J/\psi$ formation time effects, but they depend strongly for heavy target nuclei on the charmonium--nucleon
absorption cross section. We see yet that for the incoming photon energy of 11 GeV
the value of the absolute $J/\psi$ meson yield is of the order of 20--140 nb for targets
heavier than the Al target in employed four scenarios for the cross section $\sigma_{J/\psi N}$.
This value is large enough to be measured in the future CEBAF experiments. Therefore, one can conclude
that, as before in [54], the observation of the A dependence, like that just considered, offers
the possibility to determine the genuine $J/\psi N$ absorption cross section. The comparison of the results of
calculations of the total cross section of $J/\psi$ photoproduction from primary ${\gamma}N \to {J/\psi}N$ reaction channels for $\sigma_{J/\psi N}=14$ mb
with and without accounting for target nucleon binding and Fermi motion indicates that the neglecting
of the nuclear effects on the struck target nucleon leads to an enhancement of the ${J/\psi}$ yield
by about a factor of 1.2. According to equations (6) and (15), this is due to the fact that the averaged over
the struck target nucleon binding and Fermi motion elementary cross section (8) at $E_{\gamma}=11$ GeV
is reduced by the same factor compared to free one (11) calculated at above energy.

In figure 3 our predictions for the transparency ratio $T_A$, defined by equations (17), (18) and calculated
using the results presented in figure 2, as a function of the nuclear mass number A for initial photon
energy of 11 GeV are given. It can be seen that, as in the preceding case as well as in the case of
near-threshold charmonium production in $pA$ collisions considered in [54],
there are no differences between the results obtained by adopting different $J/\psi$ formation lengths under consideration. On the other hand, we may observe in this
figure the experimentally separated differences ({$\sim$}15--20\%) between all calculations corresponding to different
options for the $J/\psi N$ absorption cross section only for targets heavier than the Nb target, where they are
less than {$\sim$}10\%. The above means that also this observable can be used for determining the cross section
$\sigma_{J/\psi N}$ from the future CEBAF photoproduction experiments using the heavy targets like Pb.
Looking at this figure,
one can see yet that the difference between calculations for $\sigma_{J/\psi N}=14$ mb
with and without accounting for target nucleon binding and Fermi motion (between two lower curves) is only of the
order of 4\%. This indicates clearly that indeed the simple formula (18) is well suited for calculation the
transparency ratio $T_A$ at above threshold energies. It is also nicely seen that this quantity is practically
constant at the level of about 0.96 in the case of calculations without $J/\psi$ absorption in nuclear medium
(upper curve in figure 3). In this case, it is determined, in line with equations (7) and (17), by the ratio of
the averaged cross sections $<\sigma_{{\gamma}N \to {J/\psi}N}(E_{\gamma}=11~{\rm GeV})>_A$ and
$<\sigma_{{\gamma}N \to {J/\psi}N}(E_{\gamma}=11~{\rm GeV})>_{\rm C}$, which is approximately equal to
0.96 as our calculations showed.
%%%%%%%%%%%%%%%%%%%%%%%%%%%%%%%%%%%%%%%%%%%%%%%%%%%%%%%%%%%
\begin{figure}[!h]
\begin{center}
\includegraphics[width=12.0cm]{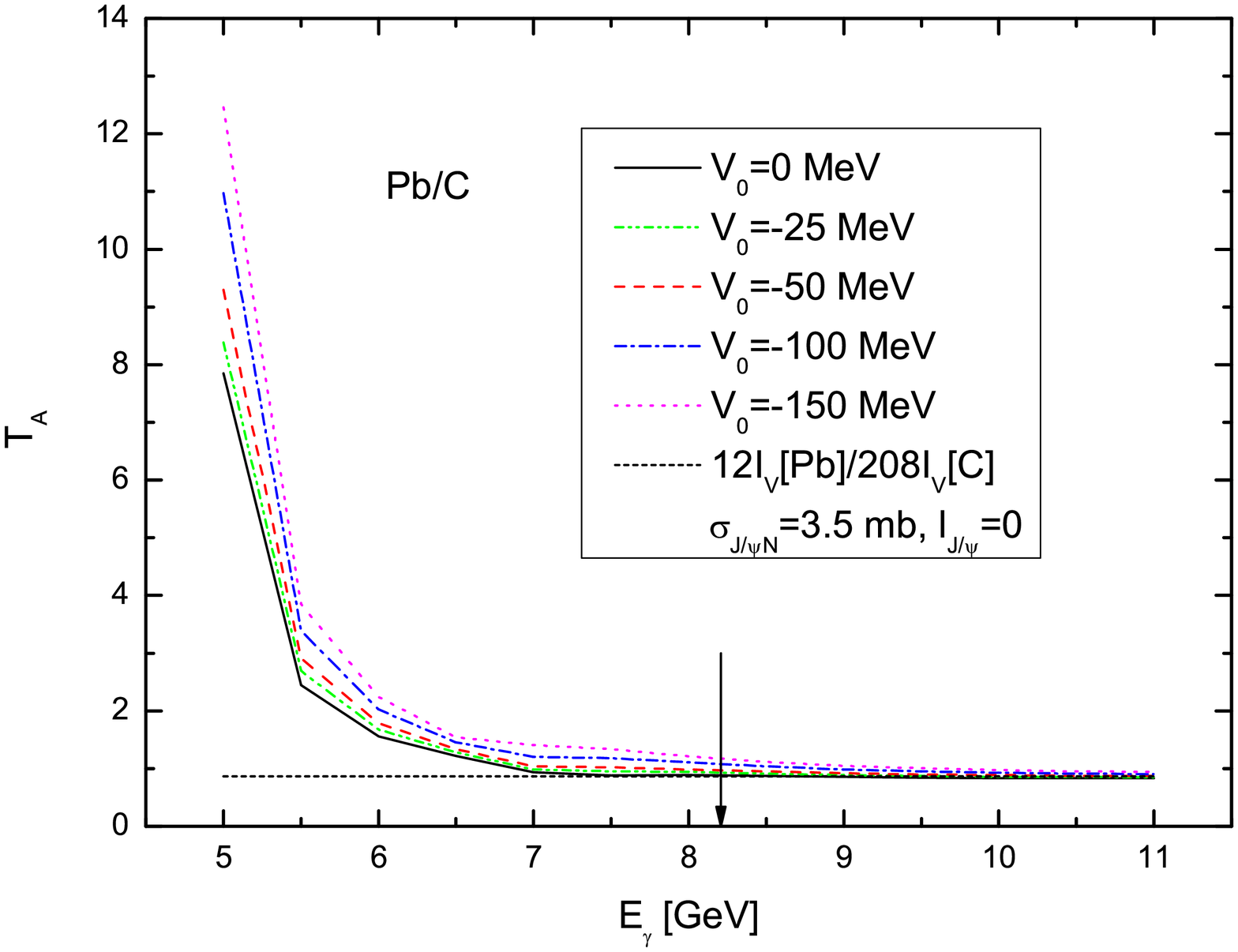}
\vspace*{-2mm} \caption{(color online) Transparency ratio $T_A$ for $J/\psi$ mesons
from primary ${\gamma}N \to {J/\psi}N$ reactions proceeding on an off-shell target nucleons
as a function of the incident photon energy for combination Pb/C. The curves are calculations by formula (17)
for the nucleus with a diffuse boundary for $\sigma_{J/\psi N}=3.5$ mb and $l_{J/\psi}=0$
with an in-medium $J/\psi$ mass shift depicted in the inset. The short-dashed straight line is
the calculation for the nucleus with a diffuse boundary by the expression (18) also for
$\sigma_{J/\psi N}=3.5$ mb and $l_{J/\psi}=0$. The arrow indicates the threshold energy
for $J/\psi$ photoproduction on a free nucleon.}
\label{void}
\end{center}
\end{figure}
%%%%%%%%%%%%%%%%%%%%%%%%%%%%%%%%%%%%%%%%%%%%%%%%%%%%%%%%%%%%%%%%%%%%%%%%%%%%%%%%%%%%%%%%%%%%%
%%%%%%%%%%%%%%%%%%%%%%%%%%%%%%%%%%%%%%%%%%%%%%%%%%%%%%%%%%%
\begin{figure}[!h]
\begin{center}
\includegraphics[width=12.0cm]{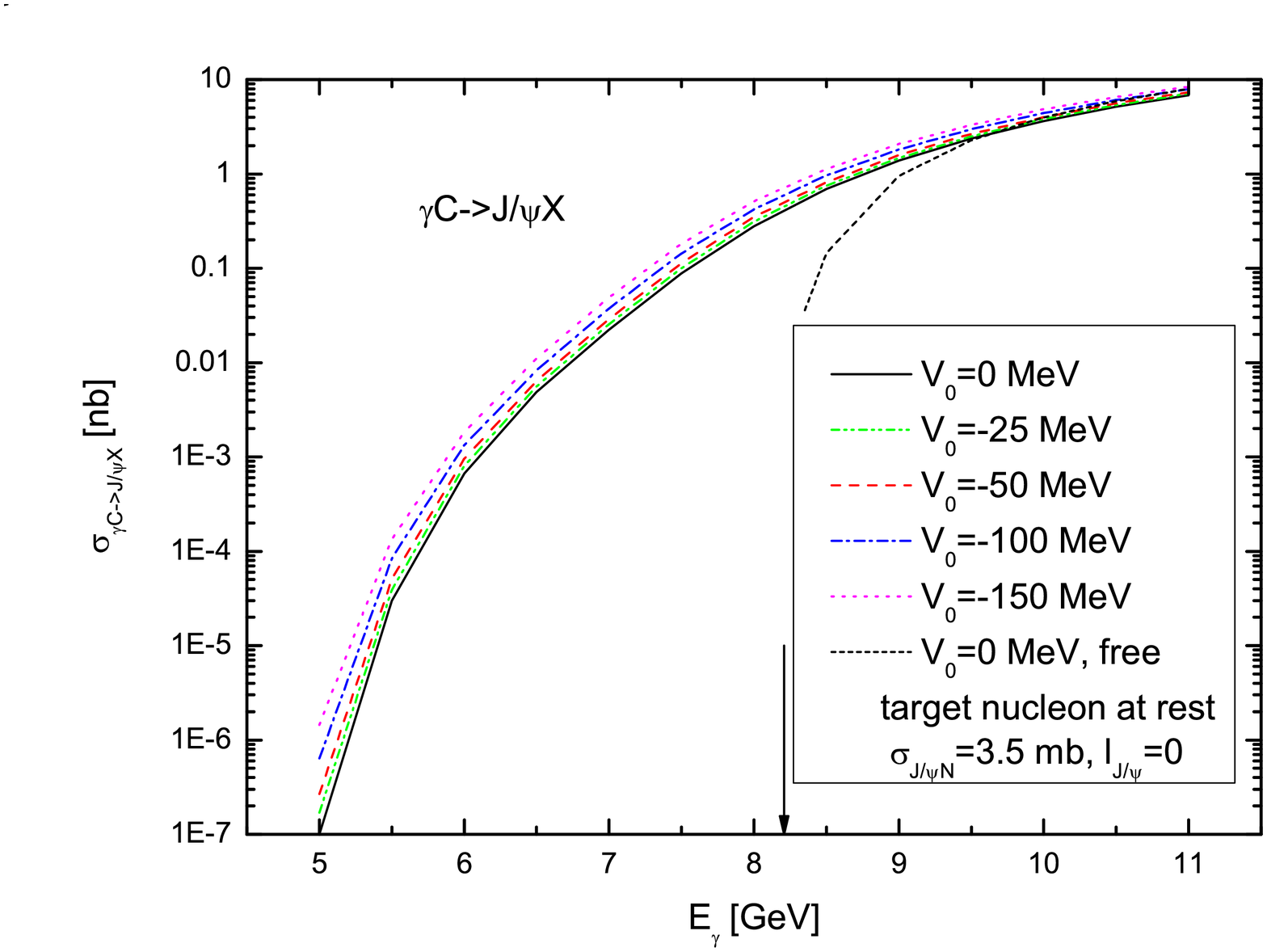}
\vspace*{-2mm} \caption{(color online) Excitation function for production of $J/\psi$
mesons off $^{12}$C from primary ${\gamma}N \to {J/\psi}N$ reactions proceeding on an
off-shell target nucleons and on a free ones being at rest.
The curves are calculations for $\sigma_{J/\psi N}=3.5$ mb and $l_{J/\psi}=0$ with
an in-medium $J/\psi$ mass shift depicted in the inset. The arrow indicates the threshold energy
for $J/\psi$ photoproduction on a free nucleon.}
\label{void}
\end{center}
\end{figure}
%%%%%%%%%%%%%%%%%%%%%%%%%%%%%%%%%%%%%%%%%%%%%%%%%%%%%%%%%%%%%%%%%%%%%%%%%%%%%%%%%%%%%%%%%%%%%
%%%%%%%%%%%%%%%%%%%%%%%%%%%%%%%%%%%%%%%%%%%%%%%%%%%%%%%%%%%
\begin{figure}[!h]
\begin{center}
\includegraphics[width=12.0cm]{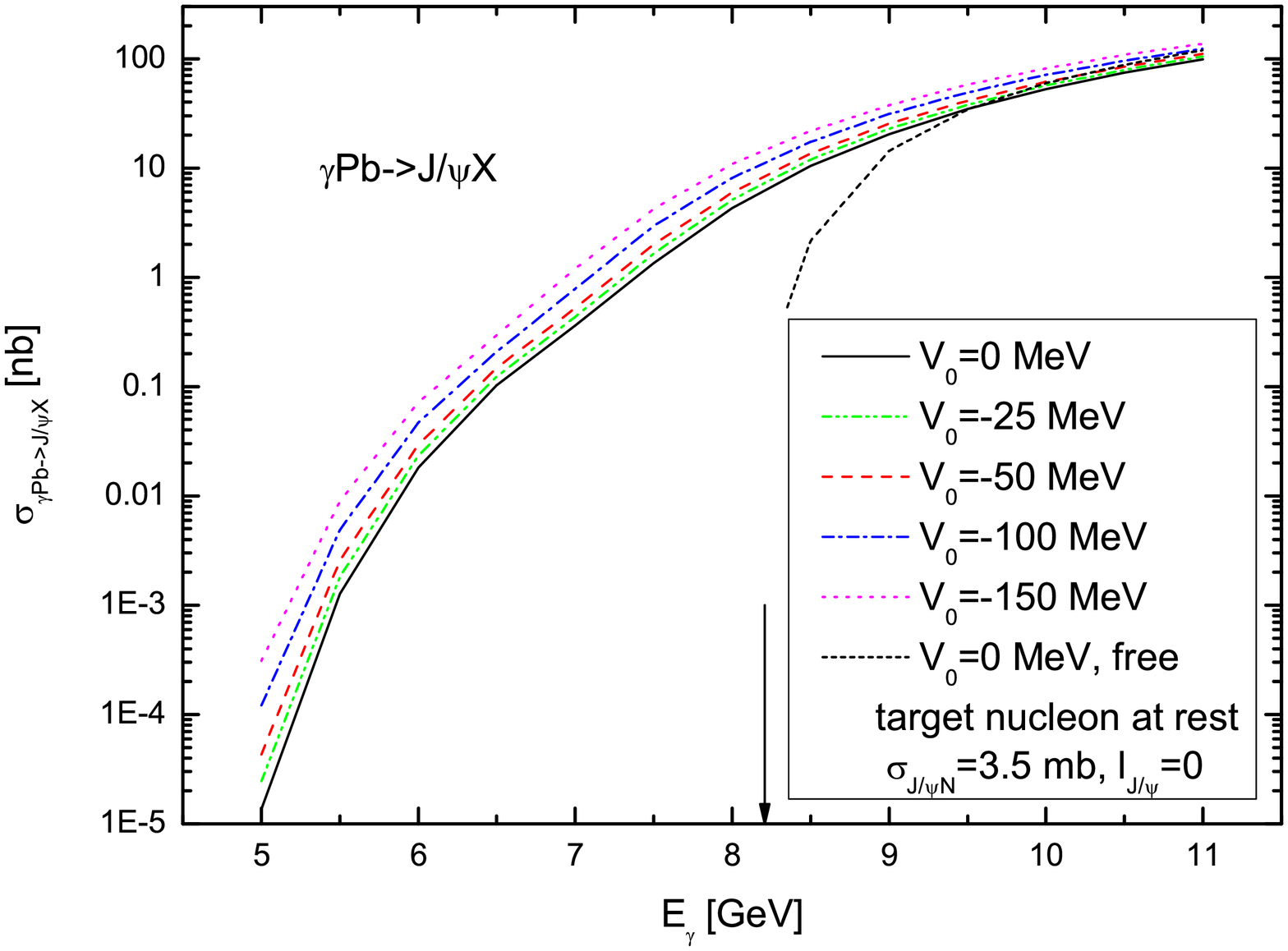}
\vspace*{-2mm} \caption{(color online) The same as in figure 5, but for the $^{208}$Pb target nucleus.}
\label{void}
\end{center}
\end{figure}
%%%%%%%%%%%%%%%%%%%%%%%%%%%%%%%%%%%%%%%%%%%%%%%%%%%%%%%%%%%%%%%%%%%%%%%%%%%%%%%%%%%%%%%%%%%%%
%%%%%%%%%%%%%%%%%%%%%%%%%%%%%%%%%%%%%%%%%%%%%%%%%%%%%%%%%%%
\begin{figure}[!h]
\begin{center}
\includegraphics[width=12.0cm]{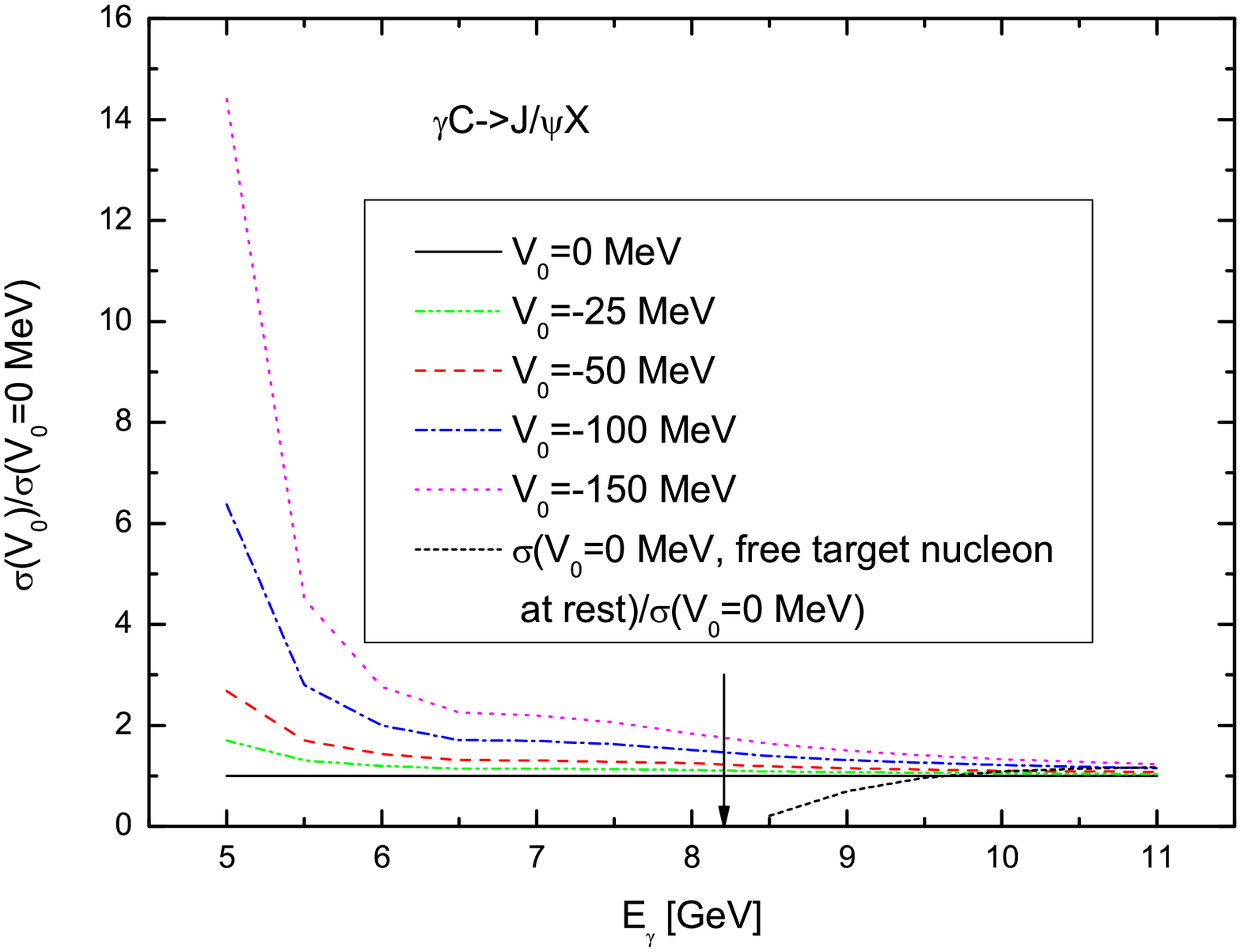}
\vspace*{-2mm} \caption{(color online) Ratio between the $J/\psi$ production cross sections on $^{12}$C
with the $J/\psi$ mass shift, shown in figure 5, and the cross section without this shift, calculated for
an off-shell target nucleons, as a function of photon energy.
The arrow indicates the threshold for the reaction ${\gamma}N \to {J/\psi}N$ occuring on a free nucleon.}
\label{void}
\end{center}
\end{figure}
%%%%%%%%%%%%%%%%%%%%%%%%%%%%%%%%%%%%%%%%%%%%%%%%%%%%%%%%%%%%%%%%%%%%%%%%%%%%%%%%%%%%%%%%%%%%%
%%%%%%%%%%%%%%%%%%%%%%%%%%%%%%%%%%%%%%%%%%%%%%%%%%%%%%%%%%%
\begin{figure}[!h]
\begin{center}
\includegraphics[width=12.0cm]{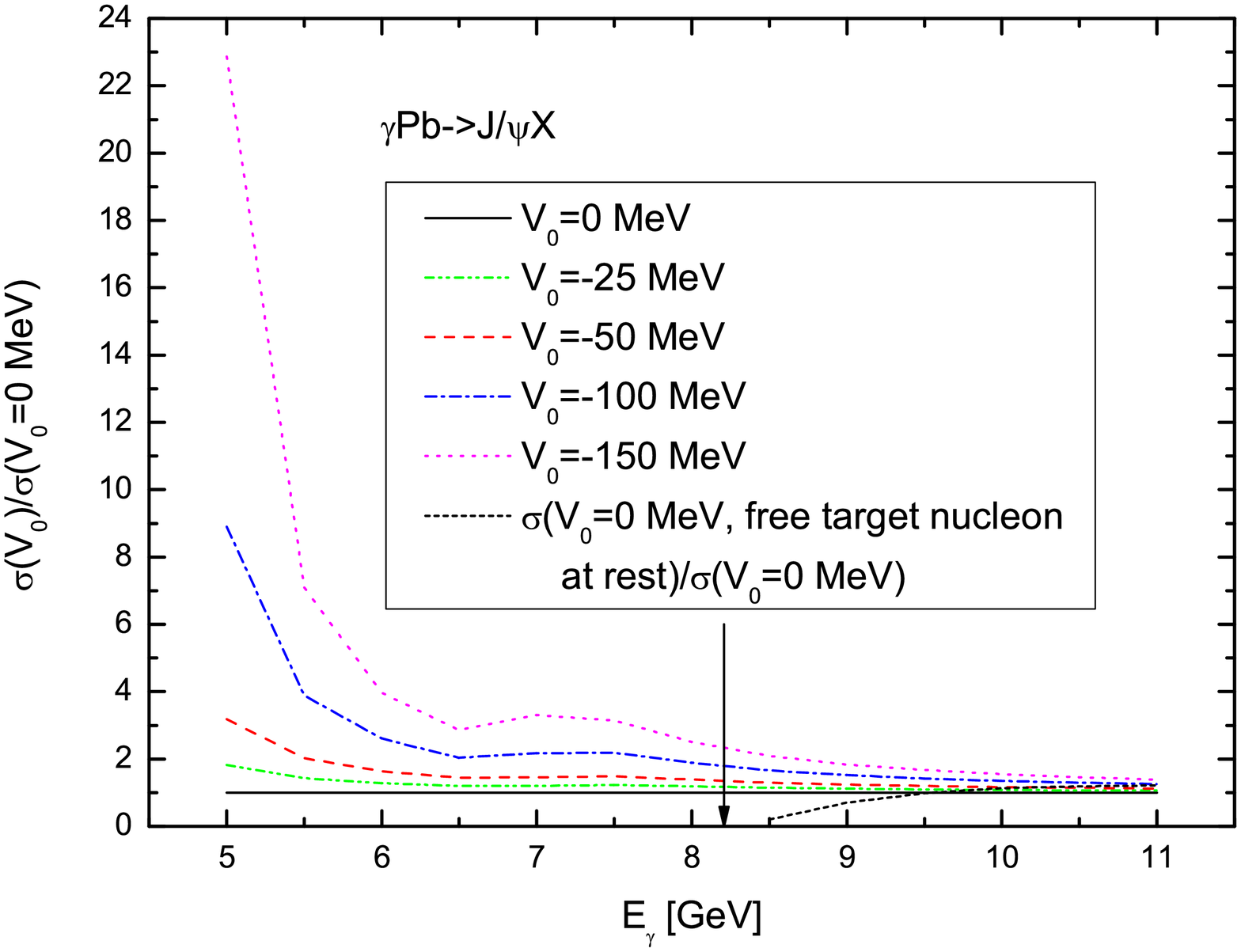}
\vspace*{-2mm} \caption{(color online) Ratio between the $J/\psi$ production cross sections on $^{208}$Pb
with the $J/\psi$ mass shift, shown in figure 6, and the cross section without this shift, calculated for
an off-shell target nucleons, as a function of photon energy.
The arrow indicates the threshold for the reaction ${\gamma}N \to {J/\psi}N$ occuring on a free nucleon.}
\label{void}
\end{center}
\end{figure}
%%%%%%%%%%%%%%%%%%%%%%%%%%%%%%%%%%%%%%%%%%%%%%%%%%%%%%%%%%%%%%%%%%%%%%%%%%%%%%%%%%%%%%%%%%%%%

   Figure 4 presents transparency ratio $T_A$ for $J/\psi$ mesons from primary ${\gamma}N \to {J/\psi}N$
reactions as a function of the initial photon energy for Pb/C combination. It was calculated on the basis
of equation (17) for $\sigma_{J/\psi N}=3.5$ mb, $l_{J/\psi}=0$ and for five employed options for the $J/\psi$
in-medium mass shift. The results of calculations for this quantity by the simple formula (18) for the same
values of $\sigma_{J/\psi N}$ and  $l_{J/\psi}$ are given here as well. It can be seen that the difference
between the various options is very small at above threshold incident photon energies $\sim$ 10--11 GeV, while
the highest sensitivity to the $J/\psi$ in-medium mass shift is found, as is expected, in the far subthreshold
region ($E_{\gamma} \sim$5--7 GeV). Here, the transparency ratio is enhanced by a factor of about 1.05 when going
from $J/\psi$ mass shift at normal nuclear matter density $V_0=0$ MeV to $V_0=-25$ MeV as well as by about the
same factor of 1.1 when going from $V_0=-25$ MeV to $V_0=-50$ MeV as when going from $V_0=-50$ MeV to $V_0=-100$ MeV
and again by about the same factor of 1.1 when going from $V_0=-100$ MeV to $V_0=-150$ MeV. On the other hand,
it is enhanced here by a factor of about 1.2 when going from $V_0=0$ MeV to $V_0=-50$ MeV as well as by about a factor
of 1.3 when going from $V_0=-50$ MeV to $V_0=-150$ MeV. It is apparent that the latter case opens an opportunity
to distinguish experimentally at least between possible zero, relatively weak ($\sim$-50 MeV) and strong
($\sim$-150 MeV) charmonium mass shifts in cold nuclear matter. But smaller mass shifts, closer to the theory
expectations, will probably not accessible by measuring the transparency ratio at subthreshold photon energies.
By looking at this figure, one can also see that the simple expression (18) describes the considered transparency
ratio quite well at incident photon energies of 7--11 GeV (compare short-dashed and solid curves), while it fails
to do that at lower beam energies.

    Now, we consider the excitation functions for production of $J/\psi$ mesons off
$^{12}$C and $^{208}$Pb target nuclei. As in [54], they were calculated on the basis of equation (6) for
$\sigma_{J/\psi N}=3.5$ mb, $l_{J/\psi}=0$ and for five adopted
scenarios for the $J/\psi$ in-medium mass shift as well as in line with formula (15) for free target nucleon
being at rest with the same values of $\sigma_{J/\psi N}$ and  $l_{J/\psi}$,
and are given in figures 5 and 6. It is seen that the difference between calculations with and without
accounting for the target nucleon Fermi motion (between short-dashed and solid curves)
is small at above threshold photon energies $\sim$ 9--11 GeV, while at lower incident energies its influence
on $J/\psi$ yield is quite essential. At energies of $\sim$ 9--11 GeV
the difference between the various scenarios for in-medium $J/\psi$ mass shift is also small.
But in the far subthreshold region ($E_{\gamma} \sim$5--7 GeV) there are well separated predictions for considered scenarios for the $J/\psi$ in-medium mass shift. The values of the total charmonium production
cross sections in this region are very small (in the range of 0.1--1000 pb), but one might expect to measure
their in the future CEBAF experiments as well. Therefore, these measurements might help to get definite information about this shift.

   To get a better impression of the size of the effect of $J/\psi$ meson in-medium mass shift on its yield
in ${\gamma}{\rm C} \to {J/\psi}X$ and  ${\gamma}{\rm Pb} \to {J/\psi}X$ reactions, in figures 7 and 8 we show
the ratios between the cross sections with mass shift and the cross sections with $V_0=0$ MeV, calculated using
the results presented in figures 5 and 6, respectively. It can be seen that the possibility of studying the
${J/\psi}$ mass shift from the excitation function measurements is feasible only in the range of mass shifts
of -50 MeV and more at far subthreshold beam energies ($E_{\gamma} \sim$5--7 GeV), while smaller mass shifts
will be not accessible by these measurements.
This is consistent with our previous findings of figure 4 and of [54].

   Taking into account the above considerations, we come to the conclusion that such observables as
the absolute and relative (transparency ratio) $J/\psi$ meson yields from ${\gamma}A$ interactions
can be useful at photon beam energies above the kinematic threshold to help determine the genuine $J/\psi N$
absorption cross section. Having this cross section fixed, the excitation function measurements at incident photon
energies below the production threshold on the free nucleon, where the effect of the charmonium mass shift in
cold nuclear matter is dominant, will allow to shed light on the possible mass shifts in the range of -50 MeV
and more.

\section*{4. Conclusions}

\hspace{1.5cm} In this paper we have calculated the A dependence of the absolute
and relative (transparency ratio) cross sections for $J/\psi$ production from ${\gamma}A$ collisions
at 11 GeV beam energy by considering incoherent direct photon--nucleon
charmonium production processes in the framework of a nuclear spectral function approach, which accounts for
the struck target nucleon momentum and removal energy distribution, elementary cross section for photon--nucleon
reaction channel close to threshold as well as different scenarios for the genuine $J/\psi N$ absorption cross section
and its formation length. Also we have calculated the absolute and relative
excitation functions for $J/\psi$ production off $^{12}$C and $^{208}$Pb target nuclei at
near-threshold incident photon energies of 5--11 GeV.
It was found that the absolute and relative $J/\psi$ yields at incident energies of 9--11 GeV,
on the one hand, is practically not influenced by formation length and mass shift effects
and, on the other hand, it is
appreciably sensitive to the charmonium--nucleon absorption cross section. This gives a nice opportunity to
determine it experimentally studying the above observables in this energy domain.
It was also shown that the absolute and relative excitation functions for
$J/\psi$ production off nuclei is well sensitive to the possible $J/\psi$ in-medium mass shifts
in the range of -50 MeV and more at subthreshold beam energies, and  this offers the possibility to investigate
such shifts via $J/\psi$ production on light and heavy target nuclei at these energies.
\\

%%%%%%%%%%%%%%%%%%%%%%%%%%%%%%%%%%%%%%%%%%%%%%%%%%%%%%%%%%%%%%%%
\end{document}